\documentclass[twocolumn,aps,prl,groupedaddress]{revtex4}
\usepackage{amsmath}
\usepackage{graphicx}

\begin{document}

\title{Equilibrium free energies from non-equilibrium metadynamics}
\author{Giovanni Bussi}
\email{gbussi@phys.chem.ethz.ch}
\author{Alessandro Laio}
\author{Michele Parrinello}
\affiliation{Computational Science, Department of Chemistry and Applied Biosciences,
Eidgen\"ossische Technische Hochschule Z\"urich,
c/o USI Campus, Via Buffi 13, CH-6900 Lugano, Switzerland}

\begin{abstract}
In this paper we propose a new formalism to map history-dependent
metadynamics in a Markovian process. We apply this formalism to a model
Langevin dynamics and determine the equilibrium distribution of a collection
of simulations. We demonstrate that the reconstructed free energy is an
unbiased estimate of the underlying free energy and analytically derive an
expression for the error.
The present results can be applied to other history-dependent stochastic
processes such as Wang-Landau sampling.
\end{abstract}

\maketitle











In recent years increasing attention has been paid to the possibility of
studying equilibrium thermodynamical processes by means of non-equilibrium
processes \cite{grub+96sci,jarz97prl,izra+97bj,geis-dell04jpcb,ober+05jpcb}.
A major breakthrough in this field is the work of 
Jarzynski \cite{jarz97prl} who has demonstrated that it is possible to
estimate the free energy difference between two states as a suitable average
of the work done on the system by forcing the transition in a finite
time.

More recently, two of us have introduced, on a more empirical basis, the
metadynamics method \cite{laio-parr02pnas}
in which the free energy as a
function of one or more collective variables (CVs) is obtained from a
non-equilibrium simulation. In this method, the dynamics of a system at
finite temperature is biased by a history-dependent potential constructed as
the sum of Gaussians centered on the trajectory of the CVs. After a transient
period, the free energy dependence on the CVs can be estimated as the
negative of the bias potential. 
This method is closely related to the local elevation method \cite{hube+94jcamd},
to coarse molecular dynamics \cite{gear+02cce,humm-kevr03jcp}
and to the adaptive-force bias method \cite{darv-poho01jcp}.
Moreover, as described in
Ref.~\cite{mich+04prl}, metadynamics can be viewed as a finite temperature
extension of the Wang-Landau approach \cite{wang-land01prl}, where the
density of states of a system is estimated by a Monte Carlo procedure in
which the acceptance probability of a move is modified every time a
configuration is explored. The practical validity of the metadynamics method
has been demonstrated in a number of applications to real problems
\cite{laio-parr02pnas,iann+03prl,mich+04prl,%
gerv+05jacs,ensi-klei05pnas,ogan+05nat,kost+05prl,iked+05jcp,jug+05ss,asci-sagu05jpca},
and an empirical way to evaluate the error has been suggested
in Ref.~\cite{laio+05jpcb}. Attempts at a more formal approach have
so far been frustrated by the lack of a formalism capable of
handling a non-Markovian process
\cite{footnote1}.

The problem of working with history-dependent dynamics is that
the forces (or the transition probabilities)
on the system depend explicitly on its history. Hence it is
not \emph{a priori} clear if, and in which sense, the system can reach a
stationary state under the action of these dynamics. In this Letter we
introduce a formalism that allows us to demonstrate that this is indeed the
case, at least when the evolution of the system is of the Langevin type.
We introduce a novel mapping of the history-dependent evolution into a Markovian
process in the original variable and in an auxiliary field that keeps track
of the configurations visited. Using this mapping we are able to  validate
rigorously the metadynamics method. In particular, we show that the average
over several independent simulations of the metadynamics biasing potential
is exactly equal to minus the free energy, and we obtain an explicit
expression for the standard deviation of the single realization from this
average. The same formalism can be extended to Monte
Carlo-like samplings such as Wang-Landau and, more generally, to all
stochastic processes augmented by an
history-dependent term which is an  explicit function of the system 
trajectory.

We will here consider the evolution of the CVs in the framework
of stochastic differential equations.
Dimensional reduction \cite{zwan61pr,oett98pre}
leads in general to a process with a complex
memory friction and an inertial term.
However, we have extensively checked \cite{laio+05jpcb,rait+06jpcb}
that in real systems the \emph{quantitative} behavior of metadynamics is
perfectly reproduced by the Langevin equation in its strong
friction limit. This is due to the fact that in real systems
all the relaxation times
are usually much smaller than the typical diffusion time in the CVs space,
and are therefore averaged out during a metadynamics reconstruction.
Hence, we model the CVs evolution as a Langevin-type dynamics.
For this dynamics it is possible to solve
analytically the equilibrium distribution of the system.

Under this assumption the metadynamics equation in the CVs $s$ of
a system with free energy $F(s)$ becomes 
\begin{equation}
ds=-D\frac{\partial }{\partial s}\bigg[F(s)+\int_{0}^{t}dt^{\prime
}g(s,s(t^{\prime }))\bigg]\bigg|_{s=s(t)}dt+\sqrt{2D}dW~,  \label{meta1}
\end{equation}%
where $dW$ is a Wiener noise, $D$ is the diffusion coefficient and we
measure the energies in units of temperature.
 The variable $s$ is in general
multi-dimensional and $\partial / \partial s$ indicates a vector
derivative in the multi-dimensional space of the CVs. The second term in
square brackets in Eq.~\ref{meta1} is the history-dependent potential, generated
through the kernel $g(s,s^{\prime })$.  So far $%
g(s,s^{\prime })$ has been taken to be a Gaussian
\cite{laio-parr02pnas,laio+05jpcb}
in the distance $%
|s-s^{\prime }|$ with a pre-factor related to the speed with which we wish to
reconstruct $F(s)$, but different kernels can be considered.
A stationary state can be reached if the system is confined in a region $\Omega$.
The analysis is simplified by considering reflecting conditions at
its boundaries $\partial \Omega$ and assuming that the
gradient of the free energy in the direction normal to $\partial \Omega $
vanishes. Other boundary conditions could easily be introduced at the cost
of algebraic complications.
The kernel $g(s,s^{\prime })$ is required
to satisfy the same boundary conditions as $F(s)$ for
any fixed value of $s^{\prime }$,
and  to be such that the equation   
\begin{equation}
\int ds^{\prime }g(s,s^{\prime })\varphi _{0}(s^{\prime })+F(s)=0
\label{def-phi0}
\end{equation}%
has a solution for the function $\varphi _{0}(s)$. 

In order to study the average properties of an ensemble of independent metadynamics
calculations we have to transform the stochastic description of Eq.~\ref%
{meta1} in a probabilistic description. When the stochastic evolution is
Markovian, this is done using the Fokker-Planck equation. However, Eq.~\ref%
{meta1} contains a history-dependent term (the bias potential) and is
clearly non-Markovian. In order to circumvent this problem we define a
time-dependent field $\varphi (s;t)$  
\begin{equation}
\varphi (s;t)=\int_{0}^{t}dt\delta (s-s(t))+\varphi _{0}(s)
\label{histogram}
\end{equation}%
which is made up of two terms: the histogram of the positions
already visited by the system and a time-independent gauge term
$\varphi _{0}(s)$ defined by Eq.~\ref{def-phi0},
introduced to simplify the resulting equations.
With this choice of the gauge it is implicitly assumed that
the initial conditions are $\varphi (s;0)=\varphi
_{0}(s)$. In terms of the  variables $s(t)$ and $\varphi (s;t)$ the stochastic
process in Eq.~\ref{meta1} can be rewritten in the simple form
\begin{subequations}
\label{meta_markovian}
\begin{align}
& ds(t)=-D\int ds^{\prime }\frac{\partial g(s,s^{\prime })}{\partial s}%
\varphi (s^{\prime };t)\bigg|_{s=s(t)}dt+\sqrt{2D}dW  \label{meta_markovian1}
\\
& d\varphi (s;t)=\delta (s-s(t))dt
\label{meta_markovian2}
\end{align}
\end{subequations}
as can be verified by direct substitution. This is the crucial step that
allows the non-Markovian evolution of a single dynamic variable $%
s(t)$ in Eq.~\ref{meta1} to be replaced with a Markovian evolution
for the extended  set of
variables which includes $s(t)$ and the field $\varphi (s;t)$. 
In fact, the state of the system at time $%
t+dt$, $[ s\left( t+dt\right) ,\varphi \left( s;t+dt\right) ]$
depends only on the state of the system at time $t,[ s\left( t\right)
,\varphi \left( s;t\right) ]$.  
The information related to the underlying free energy $F(s)$
has disappeared from the equation of motion but is still present through the
initial condition for $\varphi (s;t)$, see Eq.~\ref{histogram}.

Using the Markovian property it is possible to
analyze in a rigorous manner the behavior of Eq.~\ref{meta_markovian}. In
particular, by using standard techniques \cite{gard03book},
it is possible to write  a generalized
Fokker-Plank equation and study its asymptotic behavior for large $t$.
 We consider an
ensemble of independent metadynamics runs, and define an ensemble density.
Since our dynamic variables are the position of the walker $s$ and the field 
$\varphi (s)$, the probability density will be a function of $s$ and a
functional of $\varphi $. We denote this probability as $P(\{\varphi\},s;t)$.
The Fokker-Planck equation for $P(\{\varphi\},s;t)$ is
\begin{widetext}
\begin{equation}
\label{fokkerplanck}
  \frac{\partial P(\{\varphi\},s;t)}{\partial t} =
- \frac{\delta P(\{\varphi\},s;t)}{\delta{\varphi(s)}}
+
  D P(\{\varphi\},s;t) \int d s'
  \frac{\partial^2 g(s,s')}{\partial s^2} \varphi(s')
+ D \frac{\partial P(\{\varphi\},s;t)}{\partial s} \cdot
  \int d s'
 \frac{\partial g(s,s')}{\partial s}\varphi(s')
+ D \frac{\partial^2 P(\{\varphi\},s;t)}{\partial s^2}~.
\end{equation}
\end{widetext} Here, if the dimensionality of the system is higher than 1, 
a trace is implied and the second derivative is in fact a Laplacian.
The probabilistic description in Eq.~\ref{fokkerplanck} is completely
equivalent to the coupled stochastic Eqs.~\ref{meta_markovian1}
and~\ref{meta_markovian2}.

Equation \ref{fokkerplanck} is our main result and describes the evolution
of an ensemble of metadynamics runs. 
We would like to stress that this result has far more general relevance
than its application to the Langevin model in Eq.~\ref{meta1}. In fact,
our formalism would allow mapping the metadynamics equations 
into a Markovian form also
before performing the dimensional reduction.
For example it could be applied to
the Hamilton equations of motion in the canonical coordinates of the system,
$p$ and $q$, augmented with a Langevin thermostat in order to
impose the temperature. This would lead to a set of Markovian
equations in the original coordinates of the system and in
the field $\varphi (s;t)$, and to a Fokker-Plank equation
in a probability $P(\{\varphi\},p,q;t)$.

We now look for the limiting distribution of Eq.~\ref{fokkerplanck}
when $t\rightarrow\infty$, namely the probability density $\bar{P}$
which satisfies $\frac{\partial \bar{P}(\{\varphi\},s;t)}{\partial t}=0$.
Remarkably, this solution is independent on $s$ and is
\begin{equation}
\bar{P}(\{\varphi\})=C\exp \bigg(\frac{D}{2}\int dsds^{\prime }\varphi (s)\frac{%
\partial ^{2}g(s,s^{\prime })}{\partial s^{2}}\varphi (s^{\prime })\bigg)~,
\label{equilibrium}
\end{equation}
as can be verified by direct substitution.
Strictly speaking not all initial conditions might flow
to this solution. However, extensive numerical experimentation
has shown this not to be the case in practical applications.
$C$ is a normalization constant, and the kernel
$\frac{\partial ^{2}g(s,s^{\prime })}{\partial s^{2}}$ is
assumed to be symmetric and negative definite.

These kernel properties are better discussed through a change of basis.
The most general form for $g$ with the correct properties is
\begin{equation}
\label{gkernel}
g(s,s^{\prime })=\sum_ka_{k}(s)g_ka_{k}(s^{\prime }),~\text{with}%
~g_{k}>0~,
\end{equation}
where the $a_{k}(s)$ are the eigenfunctions of the Laplacian operator on $\Omega$.
In the one-dimensional case the label $k$ is a positive or null integer
and the basis functions are
$a_0(s)=\sqrt{1/S}$ and
$a_k(s)=\sqrt{2/S} \cos(\frac{\pi k s}{S})$ for $k\ne 0$.
For a cubic $d$ dimensional domain with side $S$,
the eigenfunctions can be factorized and the label $k$
is a $d$ dimensional vector of positive or null integers.
For $g_k$ in Eq.~\ref{gkernel} we can chose the Fourier transform of a general
radial function.
If this is a Gaussian with standard deviation $\delta s$,
$g_k\propto \exp(-\frac{\pi^2k^2\delta s^2}{2S^2})$,
where $k^2$ is the square norm of the vector $k$.
It is easily verified that the form in Eq.~\ref{gkernel} for
the kernel
is equivalent to adding not only a
Gaussian centered on the actual value of the CVs,
but also reflected Gaussians that are positioned at an equal distance on the
other side of the boundaries.
This form of the kernel is slightly different from the one introduced in
Refs.~\cite{laio-parr02pnas,laio+05jpcb}, 
but
eliminates the systematic errors close to the boundaries that are
observed using the simple Gaussians \cite{mich+04prl}
and produces a reconstructed free energy that is reliable
everywhere on $\Omega$.
This has been extensively checked on a variety of model systems.

Equation \ref{equilibrium} expresses the probability of obtaining
a given field $\varphi$ at the end of a metadynamics simulation.
Since the negative of the biasing potential
is used to estimate the free energy,
we define the error $\epsilon(s)$ as the sum of
the exact underlying free energy and the biasing potential.
Using Equations \ref{def-phi0} and \ref{histogram} we find that the error is
linearly related to the field
$\varphi$ through
\begin{equation}
 \epsilon(s)=\int ds^{\prime}g(s,s^{\prime})\varphi(s^{\prime})~.
\end{equation}
Equation \ref{equilibrium} implies that for a specific realization
of a metadynamics process the probability of finding large errors
in the estimation of the free energy is small.
Using Eq.~\ref{equilibrium} we can explicitly calculate the
expected average error of a series of runs.
Since the distribution is a Gaussian with respect to $\varphi$,
the expectation value of this field is vanishing.
The error $\epsilon(s)$ is linear in the field $\varphi(s)$,
 and consequently also its expectation value is vanishing:
\begin{equation}
\langle \epsilon(s) \rangle = 0~.
\end{equation}
Thus,
we proved that the average of the biasing potential over
a series of metadynamics runs provides an
unbiased estimate for the underlying free energy.

Using Eq.~\ref{equilibrium} we can also
address the problem of the accuracy, determining the expected
quadratic deviation $\langle \epsilon^2(s) \rangle$ of a single metadynamics
run from the average.
This expectation value can be easily calculated on the
basis of the eigenfunctions of the Laplacian:
\begin{equation}
\langle \epsilon^2(s) \rangle = \frac{S^2}{D} \sum_{k\neq 0} \frac{g_ka_k^2(s)}{\pi^2k^2}~.
\end{equation}
The average value of the error in the domain $\Omega$ is
\begin{equation}
\label{error}
\langle \epsilon^2 \rangle = \frac{1}{S^d} \int ds \langle \epsilon^2(s)
\rangle = \frac{S^2}{DS^d}\sum_{k\neq 0} \frac{g_k}{\pi^2k^2}~.
\end{equation}
A formal generalization of these expressions to domains of
a different shape is straightforward.

\begin{figure}
\includegraphics[clip,width=0.45\textwidth]{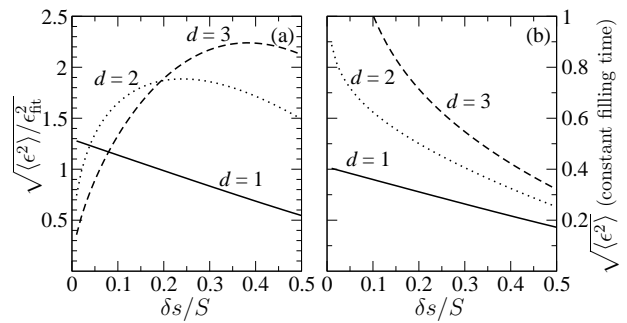}
\caption{\label{figura}
(a)
Ratio between the present expression for the error
and the formula obtained by fitting results for 
$d=1$ and $2$ in Ref.~\cite{laio+05jpcb}.
Since in the $d=3$ case the constant $C(d)$ was not estimated,
we assume here $C(3)=C(2)$.
(b)
Plot of the square root of the sum in Eq.~\ref{error_gaussian},
indicating the dependence of the error
on $\delta s/S$ for a fixed filling time
(see text for details).
}
\end{figure}
So far the results are quite general and can be used to
optimize the simulation parameters.
Since all the metadynamics simulations carried out so far
are based on a Gaussian kernel, it is
useful to specialize our results to this case
and compare the error estimate in Eq.~\ref{error} with
the empirical expression given in Ref.~\cite{laio+05jpcb}.
In order to facilitate the comparison we shall use the same
conventions as in Ref.~\cite{laio+05jpcb}, 
that is to say
reintroducing standard energy units we write
in $k$ space 
$g_k=\frac{w(\sqrt{2\pi}\delta s)^d}%
{\tau_G}\exp(-\frac{\pi^2k^2\delta s^2}{2S^2})$,
where the energy $w$ is the Gaussian strength and $1/\tau_G$
is the frequency at which the Gaussians are added to the bias.
In this case Eq.~\ref{error} gives
\begin{equation}
\label{error_gaussian}
  \langle \epsilon^2 \rangle =
\frac{S^2w}{\beta D\tau_g}\bigg(\frac{\delta s\sqrt{2\pi}}{S}\bigg)^d
\sum_{k\neq 0} \frac{\exp(-\frac{\pi^2k^2\delta s^2}{2S^2})}{\pi^2k^2}~.
\end{equation}
This is to be compared with the empirical expression
$\epsilon_{\text{fit}}
= C(d)\sqrt{\frac{S^2w}{\beta D\tau_G}\frac{\delta s}{S}}$,
where $C(d)$ is a constant depending on the dimensionality,
namely $0.5$ for $d=1$ and $0.3$ for $d=2$.
Since the dependence on $\beta$, $D$, $S$, $w$ and $\tau_G$ is
identical, we compare here the error as a function
of $\delta s/S$.
In Fig.~\ref{figura}(a) it can be seen that
the empirical expression works quite well, in spite
of the fact that it was fitted on a very small range
of Gaussian widths, namely $\delta s/S\in[0.003,0.05]$,
and that the total error was averaged discarding
the region near the boundaries.

In Ref.~\cite{laio+05jpcb} we also noticed that the
total simulation time required to fill the entire domain
is proportional to
$\frac{\tau_g}{w} \left(\frac{S}{\delta s}\right)^d$.
Therefore the sum in Eq.~\ref{error_gaussian} is
proportional to the square error at fixed simulation
time, and is a function of the dimensionless ratio $\delta s/S$
and of the dimensionality $d$.
As can be observed in Fig.~\ref{figura}(b), this quantity
is a decreasing function of the Gaussian width.
Thus, to optimize the accuracy of a metadynamics calculation,
the width has to be chosen as large as possible,
the only limit being the resolution needed to describe the underlying
free energy.
The Wang-Landau sampling as formulated in Ref.~\cite{wang-land01prl}
can be viewed as a history-dependent stochastic sampling
in which the kernel is a Kronecker delta. The present
analysis suggests that the use of a smoother kernel might
be advantageous.

As a final remark, we notice that a similar analysis
can be carried out also in the multiple-walkers
extension of metadynamics \cite{rait+06jpcb},
in which $N_w$ independent processes contribute to the
reconstructed free energy.
Equation~\ref{meta_markovian2} is generalized as
\begin{equation}
d\varphi (s;t)=\sum_{i=1}^{N_w} \delta (s-s_i(t))dt~,
\end{equation}
where $s_i(t)$ is the trajectory of the walker $i$.
It is straightforward to show that the asymptotic
probability distribution of the system is also in
this case independent
of $s_i$ and given by Eq.~\ref{equilibrium}.
This confirms the empirical result discussed in Ref.~\cite{rait+06jpcb}
that the error does not depend
on the number of walkers.

In conclusion,
the approach introduced in this Letter allows
a history-dependent dynamics such as
metadynamics to be mapped in a Markovian process where the estimated
free energy is treated as a dynamical variable.
We have applied this formalism to a Langevin model system.
When the proper collective variables of a reaction
are used, this model is representative of a large class
of realistic systems.
Our approach allows this stochastic dynamics to be treated
in a probabilistic manner and to search for its
equilibrium distribution.
We were able to demonstrate analytically the correctness
of metadynamics, and we obtained an explicit expression
for the error in the
estimated free energy at the end of a metadynamics simulation.
The present work is a step towards the understanding of all the
sampling methods based on adaptive biases.

The authors are very grateful to Eric Vanden-Eijnden for
carefully reading the manuscript and for several important suggestions.
We also acknowledge Cristian Micheletti, Paolo De Los Rios
and Giovanni Ciccotti for stimulating discussions.

\end{document}